\documentclass[preprint,prl,color,superscriptaddress,]{revtex4}
\usepackage[dvipdfmx]{graphicx} 
\usepackage[english,russian]{babel}
\usepackage{graphicx}
\usepackage{epsfig}
\usepackage{epsfig}
\usepackage{color,amsmath,latexsym}

\begin{document}

\title{DFT, L(S)DA, LDA+U, LDA+DMFT..., whether we do approach to a proper description of optical response for  strongly correlated systems?}
\author{A.S. Moskvin}
\affiliation{Ural Federal University, Ekaterinburg, 620083 Russia
}

\begin{abstract}
I present a critical overview of so-called "{\it ab initio}" DFT (density fuctional theory) based calculation schemes for the description of the electronic structure, energy spectrum, and optical response for strongly correlated 3$d$ oxides, in particular, crystal-field and charge transfer transitions as compared with an "old"\, cluster model that does generalize crystal-field and ligand-field theory. As a most instructive illustration of validity of numerous calculation techniques I address the prototypical 3$d$ insulator NiO predicted to be a metal in frames of a standard LDA (local density approximation) band theory. 
\end{abstract}

\maketitle

\section{Introduction}

The electronic states in strongly correlated 3$d$ oxides manifest both significant localization and dispersional features. One strategy to deal with this dilemma is to restrict oneself to small many-electron clusters embedded to a whole crystal, then creating model effective lattice Hamiltonians whose spectra may reasonably well represent the energy and dispersion of the important excitations of the full problem. Despite some shortcomings the method did provide a clear physical picture of the complex electronic structure and the energy spectrum, as well as the possibility of a quantitative modeling. 

However, last decades the condensed matter community faced an expanding flurry of papers with the so called {\it ab initio} calculations of electronic structure and physical properties for strongly correlated systems such as 3$d$ compounds  based on density functional theory (DFT). The modern formulation of the DFT originated in the work of Hohenberg and Kohn\,\cite{HK}, on which based the other classic work in this field by Kohn and Sham\,\cite{KS}. The Kohn-Sham equation, has become a basic mathematical model of much of present-day methods for treating electrons in atoms, molecules, condensed matter, solid surfaces, nanomaterials, and man-made structures\,\cite{Kryachko}. Of the top three most cited physicists in the period 1980-2010, the first (Perdew: 65 757 citations) and third (Becke: 62 581 citations) were density-functional theorists\,\cite{Becke}.

However, DFT still remains, in some sense, ill-defined: many of DFT statements were ill-posed or not rigorously proved.
Most widely used DFT computational schemes start with a "metallic-like" approaches making use of approximate energy functionals, firstly LDA (local density approximation) scheme, which are constructed as expansions around the homogeneous electron gas limit and fail quite dramatically in capturing the properties of strongly correlated systems. 
         The LDA+U and LDA+DMFT (DMFT, dynamical mean-field theory)\,\cite{LDA+U+DMFT} methods are believed to correct the inaccuracies of approximate DFT exchange correlation functionals. The main idea of these computational approaches consists in a selective description of the strongly correlated electronic states, typically, localized $d$ or $f$ orbitals, using the Hubbard model, while all the other states continue to be treated at the level of standard approximate DFT functionals. 
          At present the LDA+U and LDA+DMFT methods are addressed to be most powerful tools for the investigation of strongly correlated electronic systems, however, these preserve many shortcomings of the DFT-LDA approach. Despite many examples of a seemingly good agreement with experimental data (photoemission and inverse-photoemission spectra, magnetic moments,...) claimed by the DFT community, both the questionable starting point and many unsolved and unsoluble problems give rise to serious doubts in quantitative and even qualitative predictions made within the DFT based techniques.  
In a certain sense the cluster based calculations seem to provide a better description of the overall electronic structure of insulating 3$d$oxides and its optical response than the DFT based band structure calculations, mainly due to a clear physics and a better account for correlation effects (see, e.g., Refs.\,\cite{Eskes,Ghijsen}). 

					
					The paper is organized as follows. In Sec.II we do present a short critical overview of the DFT and the DFT based technique with a focus on the NiO oxide. Sec.III is devoted to a short overview of the cluster model approaches to a proper semiquantitative description of the optical response in strongly correlated 3$d$ oxides with a focus on the NiO oxide. A short summary is made in Sec.IV.

\section{Short overview of the DFT based technique}

\subsection{Hohenberg-Kohn-Sham DFT}

Density functional theory finds its roots in the approach which Thomas and Fermi elaborated shortly after the creation
of quantum mechanics\,\cite{A,B}. The Thomas-Fermi theory of atoms may be interpreted as a semiclassical approximation,
where the energy of a system is written as a functional of the one-particle density.

Justifying earlier attempts directed at generalizing the Thomas-Fermi theory, Hohenberg and Kohn\,\cite{HK} in 1964  advanced a theorem: "For any system of interacting particles in an external potential $v({\bf r})$, the external potential is uniquely determined (except for a constant) by the ground state density $n_0({\bf r})$", which states that the exact ground-state energy is a functional of the exact ground-state one-particle density. Unfortunately, it does not tell how to construct this functional, 
i.e., it is an existence theorem for the energy-density functional. This explains the fact of why so much effort has
been dedicated to the task of obtaining approximate functionals for the description of the ground-state properties of many-particle
systems. Contrary to wavefunction theory, where the objective is to approximate the wavefunction, in DFT we choose to make approximations for the functional.


However, DFT still remains, in some sense, ill-defined: many of the DFT statements were ill-posed or not rigorously proved. 
Indeed, the HK theorem is the constellation of two statements: (i) the mathematically rigorous HK lemma, which demonstrates that the same ground state density cannot correspond to two different potentials of an external field, and (ii) the hypothesis
of the existence of the universal density functional. However, the HK lemma cannot provide justification of the universal density functional for fermions\,\cite{Bobrov-JETP}. In other words, each external field determines a unique density, and each density determines a unique external field on the basis of the HK lemma. However, the rule for the last correspondence can be nonuniversal, as the rule
in general depends on the concrete form of the density. The existence of this nonuniversality violates the HK theorem, although the HK lemma is  believed to be undoubtedly correct\,\cite{Bobrov-JETP}. 

Furthermore, there are more serious critics. Sarry and Sarry\,\cite{Sarry} claim that the proof of the HK theorem is not correct. 
The authors do emphasize that for a strict many-particle calculation only the direct mapping: external potential $\Rightarrow$ ground state wave function $\Rightarrow$ electron density  
$$
 v({\bf r}) \Rightarrow \Psi_0({\bf r}) \Rightarrow \rho_0({\bf r})
$$
is justified while the inverse mapping 
$$
 \rho_0({\bf r}) \Rightarrow \Psi_0({\bf r}) \Rightarrow v({\bf r})
$$
claimed by the HK theorem can be validated only for single-particle self-consistent calculations.

The DFT exploits the one-to-one correspondence between the single-particle electron
density and an external potential acting upon the system and relies on the existence of a universal functional
$F[\rho ({\bf r})]$ which can be minimized in order to find the ground state energy. However,the correspondence theorem establishes the existence of the
functional only in principle, and provides no unique practical recipe for its construction. The  construction of the functional $F[\rho ({\bf r})]$ in the HK-DFT is equivalent to the  problem of finding the N-representability conditions of the reduced density matrix of order two\,\cite{Kryachko,Ludena}, the problem whose solution has not been found until now. Generally speaking the functional $F[\rho ({\bf r})]$ must be $N$-dependent, namely, $F[N,\rho ({\bf r})]$. Another important aspect, closely related to $N$-representability, is the variational character that either exact or approximate functionals $F[N,\rho ({\bf r})]$ must have in order to guarantee that the energy remains an upper bound to the
exact value.

The Kohn-Sham (KS) theory goes further in reducing the problem of calculating ground state properties of a many-electron system in a local external single-particle potential to solving Hartree-like one-electron KS equations. 
Within the framework of the HKS-DFT, the many-body problem of interacting electrons in a static external potential is cast into a tractable problem of non-interacting electrons moving in an effective potential. The latter includes the external potential and the effects of the Coulomb interactions between the electrons, i.e. the Hartree term, describing the electron-electron repulsion, and the exchange and correlation (XC) interactions, which includes all the many-body interactions. Modeling the XC interactions is the main difficulty of DFT. 
In practical calculations, the XC contribution is approximated, and the results are only as good as the approximation used. 
Actually, in HKS-DFT there exist hundreds of XC-approximations for $v^{KS}_{xc}({\bf r})$\,\cite{Kryachko}. The existence of so many
approximations, with so little guidance, makes it ever more difficult for non-specialists to separate the silver from the dross\,\cite{Burke}.
It is worth noting here that all the approximate functionals do not comply with the variational principle.

The leading approximation for density functional construction is the so called local density approximation (LDA), which is based upon exact exchange energy for a uniform electron gas and only requires the density at each point in space. So the LDA taken from assuming that the electron density for an atom, molecule, or solid is similarly homogeneous.
But molecules in LDA are typically overbound by about 1\,eV/bond, and in the late 1980s the so-called generalized gradient approximations (GGAs)  using both the density and its gradient at each point in space were elaborated whose accuracy seemed to be acceptable in chemical calculations\,\cite{Burke}. All the GGAs functionals, by definition, are corrections to the LDA, they all revert to the uniform electron gas at zero density gradient. It should be noted that the local nature of the standard approximations implies an exponential decay of the inter-site interaction, in other words, the description of  weak interactions such as long-range van der Waals interactions is well beyond any conventional DFT method\,\cite{Burke}.




The DFT calculations are quite different from the usual quantum mechanical methods where better accuracy depends on computational resources and not on limitations stemming from the method itself. The Hartree-Fock (HF) results cannot be reproduced within the framework of Kohn-Sham (KS) theory because the single-particle densities of finite systems obtained within the HF calculations are not $v$-representable, i.e., do not correspond to any ground state of a $N$ non-interacting electron systems in a local external potential\,\cite{Amusia}. For this reason, the KS theory, which finds a minimum on a different subset of all densities, can overestimate the ground state energy, as compared to the HF result.


In addition to the lack of compliance with N-representability conditions and difficulties in extending the application
of the first HK theorem to finite subspaces, there are still other problems that beset DFT. They have to do with how to
properly include symmetry (i.e., properties of all operators commuting with the Hamiltonian of a given system). For instance, translational symmetry in crystalline solids should be applied only to a full many-electron function rather than to one-electron KS orbitals!

Currently, the KS-DFT is about occupied orbitals only and is far from giving a consistent and quantitatively accurate description of open-shell spin systems, as the currently available approximate functionals show unsystematic errors in the (inaccurate) prediction of energies, geometries, and molecular properties.

Strictly speaking, the DFT is designed for description of ground rather than excited states with no good scheme for excitations. Because an excited-state density does not uniquely determine the potential, there is no general analog of HK for excited states. The standard functionals are inaccurate both for on-site crystal field and for charge transfer excitations\,\cite{Burke}. The DFT based approaches cannot provide the correct atomic limit and the term and multiplet structure, which is crucial for description of the optical response for 3$d$compounds. Although there are efforts to obtain correct results for spectroscopic properties depending on spin and orbital density this problem remains as an open one in DFT research. 
Clearly, all these difficulties stem from unsolved foundational problems in DFT and are related to fractional charges and to fractional spins. Thus, these basic unsolved issues in the HKS-DFT point toward the need for a basic understanding of foundational issues.


In other words, given these background problems, the DFT based  models should be addressed as semi-empirical approximate ones rather than  {\it ab initio}  theories. M. Levy introduced in 2010  the term DFA to define density functional approximation instead of DFT, which is believed to quite appropriately describe contemporary DFT\,\cite{Kryachko}.
In chemistry, it is  traditional to refer to standard approaches as {\it ab initio}, while DFT is regarded as empirical. Because solid-state calculations are more demanding, for many decades DFT was the only possible approach. Thus, DFT calculations are referred to as {\it ab initio} in solid-state physics and materials science\,\cite{Burke}. Proceeding with a fixed approximate functional, the DFT is called "first principles", in the sense that the user only chooses the atoms, and the computer predicts (correctly or not) all properties of the molecule or solid.



\subsection{LSDA}

Basic drawback of the spin-polarized approaches is that these start with a  local density functional in the form 
(see, e.g. Ref.\onlinecite{Sandratskii})
$$
{\bf v}({\bf r})=v_0[n({\bf r})]+\Delta v[n({\bf r}),{\bf m}({\bf r})](\bf\hat\sigma\cdot \frac{{\bf m}({\bf r})}{|{\bf m}({\bf r})|})\, ,
$$
where $n({\bf r}),{\bf m}({\bf r})$ are the electron and spin magnetic density, respectively, ${\bf \hat\sigma}$ is the Pauli matrix, that is these imply  presence of
a large fictious local {\it one-electron} spin-magnetic field $\propto (v^{\uparrow}-v^{\downarrow})$, where $v^{\uparrow ,\downarrow}$ are the on-site LSDA spin-up and spin-down potentials. Magnitude of the field is considered to be  governed by the intra-atomic Hund exchange, while its orientation does by the effective molecular, or inter-atomic exchange fields. Despite the supposedly spin nature of the field it produces an unphysically giant spin-dependent rearrangement of  the charge density that cannot be reproduced within any conventional technique operating with spin Hamiltonians. Furthermore, a  direct link with the orientation of the field makes the effect of the spin configuration onto the charge distribution to be unphysically large. However, magnetic long-range order has no significant influence on the redistribution of the charge density. The DFT-LSDA community needed many years to understand such a physically clear point. 

In general, the  LSDA method to handle a spin degree of freedom is absolutely incompatible with a conventional approach based on  the spin Hamiltonian concept. There are some intractable problems with a match making between the conventional formalism of a spin Hamiltonian and LSDA approach to the exchange and exchange-relativistic effects.
Visibly plausible numerical results for different exchange and exchange-relativistic  parameters reported in many LSDA investigations (see, e.g., Refs.\,\cite{Mazurenko})  evidence only a potential capacity of the LSDA based models for semiquantitative estimations, rather than for reliable  quantitative data.
It is worth noting that for all of these "advantageous" instances the matter concerns the handling of certain classical N\'eel-like  spin configurations (ferro-, antiferro-, spiral,...) and search for a compatibility with a mapping made with a  conventional quantum spin Hamiltonian. It's quite another matter when one addresses the search of the charge density redistribution induced by a spin configuration as, for instance, in multiferroics. In such a case the straightforward application of the LSDA scheme can lead to an unphysical overestimation of the effects or even to qualitatively incorrect results due to an unphysically strong effect of a breaking of spatial symmetry induced by a spin configuration (see, e.g. Refs.\,\cite{MF} and references therein).

 \subsection{Going beyond LSDA:LDA+U, LDA+DMFT, LDA+U+V}

It is commonly accepted now that the standard DFT-LDA(GGA) approach is insufficient to describe the electronic structure of the Mott insulators.

Apparent weaknesses of the DFT approach were exposed especially after the discovery in 1986 of the copper-oxide superconductors, as it failed to yield the fact that the parent compound La$_2$CuO$_4$ is an antiferromagnetic insulator. 
This difficult period for the DFT-LDA method as many decided was partially ended in the early and mid 1990s especially when an orbital dependent Hubbard-type U was incorporated in the exchange correlation functional of the localized 3$d$electrons within the LDA+U method, while the other electrons are still described at the LDA level\,\cite{LDA+U+DMFT}. 

Attempts to go beyond LSDA are based on the self-interaction-corrected density functional theory SIC-DFT, the LDA+U method, and the GW approximation\,\cite{LDA+U+DMFT}.  These methods represent corrections of the single-particle Kohn-Sham potential in one way or another and lead to substantial improvements over the LSDA results for the values of the energy gap and local moment. Within the SIC-DFT and LDA+U methods the occupied and unoccupied states are split by the Coulomb interaction U, whereas within the LSDA this splitting is caused by the Stoner parameter J, which is typically one order of magnitude smaller than U. Therefore, compared with the LSDA, the novel methods capture more correctly the physics of transition-metal oxides and improve the results for the energy gap and local moment significantly. 

An important drawback of the LDA+U method is that it requires U as a starting parameter. Even though several schemes for the determination of U exist, it is almost always chosen such that it reproduces the experimental value of a specific property of the electronic structure, most often the band gap.
Usually the LDA+U calculations imply account of the on-site d-d correlations with U$_{dd}$ parameter and do neglect the ligand p-p correlations though U$_{dd}$ parameter is only twice as large as U$_{pp}$ in oxides\,\cite{Eskes,Ghijsen}.
The predictive power of the novel methods crucially relies on a reliable assessment of the interactions, however, the value of the interaction parameters, such as U$_{dd}$, U$_{pp}$, depends on the choice of the downfolded model, namely, the orbitals treated in the model as well as the basis functions employed, as the screened interaction is determined by the various screening processes that are not considered in the model. Therefore a careful analysis is needed to make a proper model and choose appropriate parameters.
By fitting, one usually finds higher accuracy for systems similar to those fitted, but usually greater inaccuracies far away.

All efforts to account for the correlations beyond LDA encounter an insoluble problem of double counting (DC) of interaction terms which had just included into Kohn-Sham single-particle potential.
A well defined analytical expression for the DC potential cannot be formulated in the context of LDA+U or other technique going beyond LDA\,\cite{DC}.
How to choose the DC correction potential in a manner that is both physically sound and consistent is unknown.
 Thus, one has to resort to numerical criteria to fix the value of the DC correction. However, there is currently no universal and unambiguous expression for DC correction, and different formulations are used for different classes of materials. Moreover, different methods for fixing the double counting can drive the result from Mott insulating to almost metallic\,\cite{DC,Nekrasov-JETP}.

The LDA+DMFT approach combines band structure theory within the DFT-LDA with many-body theory as provided by dynamical mean-field theory (DMFT)\,\cite{LDA+U+DMFT}.  Within DMFT, a lattice model is mapped onto an effective impurity problem embedded in a medium which has to be determined self-consistently,  e.g., by quantum Monte-Carlo (QMC) simulations. This mapping becomes exact in the limit of infinite dimensions.

The LDA+U and LDA+DMFT methods are believed to correct the inaccuracies of approximate DFT exchange correlation functionals. The main idea of the both computational approaches consists in a selective description of the strongly correlated electronic states, typically, localized d or f orbitals, using the Hubbard model, while all the other states continue to be treated at the level of standard approximate DFT functionals. 
At present the LDA+U and LDA+DMFT methods are addressed to be most powerful tools for the investigation of strongly correlated electronic systems, however, these preserve many shortcomings of the basic DFT-LDA approach.

Current theoretical studies of electronic correlations in transition metal oxides typically only account for the local repulsion between d-electrons even if oxygen ligand p-states are an explicit part of the effective Hamiltonian. Interatomic correlations such as V$_{pd}$ between d- and (ligand) p-electrons, as well as the on-site and inter-site interaction between p-electrons (U$_{pp}$ and V$_{pp}$), are usually neglected. Strictly speaking, LDA+DMFT scheme should incorporate both U$_{pp}$,  V$_{pp}$, V$_{pd}$ and V$_{dd}$ interactions\,\cite{V}.  To this end we need a proper procedure for their calculation, however, this makes the double counting problem significantly more sophisticated. 

\subsection{NiO as a main TMO system for so-called {\it ab initio} studies}

An ongoing challenge during the last 60 years has been the development of a theoretical model that could offer an accurate description of both the electric and magnetic phenomena observed in NiO. Nickel oxide is one of the prototypical compounds that has highlighted the importance of correlation effects in transition metal oxides (TMO). However, despite several decades of studies there is still no literature consensus on the detailed electronic structure of NiO.  Although exhibiting a partially filled 3$d$band and predicted by simple band theory to be a good conductor, NiO has a relatively large band gap (about 4\,eV) that cannot be accounted for in the LDA calculations.

NiO has long been viewed as a prototype "Mott insulator"\,\cite{Brandow} with the gap formed by intersite cation-cation \emph{d}-\emph{d} charge transfer (CT) transitions, however, this view was later replaced by that of a "CT insulator" with the gap formed by  anion-cation \emph{p}-\emph{d} CT transitions\,\cite{ZSA}.


Strictly speaking, the DFT is designed for description of ground rather than excited states.  
Nevertheless research activity in the condensed matter DFT community is focused on the single-particle excitation properties of the TMOs, in particular, the photoemission spectra and energy gap.

The XPS combined with bremsstrahlung-isochromat spectroscopy (BIS) shows a gap between the top of the valence band and the bottom of the conducting band of 4.3 eV for NiO\,\cite{gap1}. Namely this value appears to be in the focus of the so-called {\it ab initio} DFT-LDA  based calculations for NiO. However, the later studies\,\cite{gap2} have shown that the exact value of this conductivity gap is subject to the band position chosen to define the highest valence and lowest conducting levels, obtaining values that range from 3.20 to 5.67\,eV (!). 
Experimental data, in particular, oxygen x-ray emission (XES) and absorption (XAS) spectra\,\cite{gap3}  point to  strong matrix element effects, that makes reliable estimates of the energy gap to be very ambiguous adventure.

The standard DFT-LDA band theory predicts NiO to be a metal. LSDA\,\cite{Terakura} predicts NiO to be an insulator (with severe underestimated  gap of 0.3\,eV) only in antiferromagnetic state (!?). The later GW\,\cite{GW} and LDA+U\,\cite{Anisimov_1993} calculations yielded the larger gap of 3.7\,eV. First LDA+DMFT calculation performed by Ren {\it et al}.\,\cite{Nekrasov_2006} yielded the value of 4.3\,eV.
The authors claimed: "The overall agreement between the calculated single-particle spectrum and the experimental data is surprisingly good". 
However, they do neglect the matrix element effect, p-d covalency, U$_{pp}$, V$_{pd}$, and V$_{dd}$, that de facto does invalidate their conclusion. Part of these effects, in particular, $p-d$ covalency was taken into account later\,\cite{Anisimov_2007}, but with a severe reinterpretation of the DOS. Again, the authors claim: "...we were able to provide a full description of the valence-band spectrum and, in particular, of the distribution of spectral weight between the lower Hubbard band and the resonant peak at the top of the valence band. 
However,  to this day the LDA+DMFT results for NiO strongly depend on the choice of the DC correction potential driving the result from Mott insulating to metallic state\,\cite{DC,Nekrasov-JETP}.



It is rather surprising how little attention has been paid to the DFT based calculations of the TMO optical properties. Lets turn to a very recent paper by  Roedl and Bechstedt\,\cite{NiO_optics} on NiO and other TMOs, whose approach is typical for DFT community.
The authors calculated the dielectric function  $\epsilon (\omega )$ for NiO within the DFT-GGA+U+$\Delta$ technique and claim:"The experimental data agree very well with the calculated curves"\, (!?). However, this seeming agreement is a result of a simple fitting when the two model parameters U and $\Delta$ are determined such (U\,=\,3.0, $\Delta$\,=\,2.0\,eV) that the best possible agreement  concerning the positions and intensities of the characteristic peaks in the experimental  spectra is obtained. In addition, the authors arrive at absolutely unphysical conclusion: "The optical absorption of NiO is dominated by intra-atomic $t_{2g} \rightarrow e_g$ transitions"\, (!?). 

Nekrasov {\it et al}.\,\cite{Nekrasov-JETP} realized the DMFT calculation of the optical conductivity for NiO. Just another correlation parameter was chosen: U\,=\,8\,eV.   The authors claim a general agreement both with optical and the X-ray experiments. In the calculations, they found that the main contribution to optical conductivity is due to intra-orbital optical transitions. Inter-orbital optical transitions give less than 5\% of the optical conductivity intensity in the frequency range used in the calculations. However, as usual they did neglect a number of important on-site and inter-site correlation parameters and all the effects due to optical matrix elements that does invalidate their conclusion. Furthermore, the DFT-LDA based schemes do not provide the correct atomic limit and the term and multiplet structure.  Hence these cannot correctly  describe both the $d$-$d$ crystal field and $p$-$d$ and $d$-$d$ charge transfer transitions. 
 However, some authors\,\cite{Misha+Sasha}  suppose that in future this problem probably can be solved within the LDA+DMFT.

Surveying these and other literature data we can argue that the conventional DFT based technique cannot provide a proper description of the optical response for strongly correlated 3$d$compounds.
As up till now, in future the optical properties of the Mott or charge transfer insulators will be considered within the framework of cluster approaches initially based on quantum-chemical calculations.

\section{Cluster model in NiO}

Cluster model approach does generalize and advance  crystal-field and ligand-field theory. The method provides a clear physical
picture of the complex electronic structure and the energy spectrum, as well as the possibility of a quantitative modelling.
In a certain sense the cluster calculations might provide a better description of the overall electronic structure of  insulating  3$d$oxides  than the band structure calculations, mainly  due to a better account for correlation effects, electron-lattice coupling, and relatively weak interactions such as spin-orbital and exchange coupling. Cluster models have proven themselves to be reliable  working models for strongly correlated systems such as transition-metal and rare-earth compounds. These have a long and distinguished history of application in optical and electron spectroscopy, magnetism, and magnetic resonance. The author with colleagues has successfully demonstrated great potential of the cluster model for description of the $p$-$d$ and $d$-$d$ charge transfer transitions and their contribution to optical and magneto-optical response  in 3$d$oxides such as ferrites, manganites, cuprates, and nickelates\,\cite{Moskvin-CT}.

Cluster models do widely use the symmetry for atomic orbitals, point group symmetry, and advanced technique such as Racah algebra and its modifications for point group symmetry\,\cite{Sugano}. From the other hand the cluster model is an actual proving-ground for various calculation technique from simple quantum chemical MO-LCAO (molecular orbital-linear-combination-of-atomic-orbitals) method to a more elaborate LDA+MLFT (MLFT, multiplet ligand-field theory)\,\cite{Haverkort} approach.



Cluster models traditionally combined  quantum chemical MO-LCAO calculations\,\cite{Sugano} based on atomic Hartree-Fock orbitals with making use parameters fitted to experiments. Several authors obtained model parameters by performing an LDA calculation for the cluster and using its Kohn-Sham MOs. 
First comprehensive  description of the electronic structure of the NiO$_6$ cluster was performed by Fujimori and Minami\,\cite{Fujimori}. Effective transfer and overlap integrals were evaluated from  LCAO parameters of NiO found by Mattheiss\,\cite{Mattheiss} by fitting their   APW energy-band results. The localized approach has been shown to successfully explain the photoemission, optical-absorption, and isochromat spectra of NiO. 
Recently, Haverkort {\it et al.}\,\cite{Haverkort} suggested a sort of generalization of conventional ligand-field model with the DFT-based calculations within a so-called "ab initio" LDA+MLFT technique.   
They start by performing a DFT calculation for the proper, infinite crystal using a modern DFT code which employs an accurate density functional and basis set [e.g., linear augmented plane waves (LAPWs)]. From the (self-consistent) DFT crystal potential they then calculate a
set of Wannier functions suitable as the single-particle basis for the cluster calculation. The authors compared the theory with experimental spectra (XAS, nonresonant IXS, photoemission spectroscopy) for SrTiO3, MnO, and NiO  and found overall satisfactory agreement, indicating that their ligand-field parameters are correct to better than 10\%. However, as in Ref.\,\cite{Fujimori} the authors  have been forced to  treat on-site correlation parameter U$_{dd}$ and orbitally averaged (spherical) $\Delta_{pd}$ parameter as adjustable ones. Comparing the novel LDA+MFLT technique with that of Fujimori and Minami\,\cite{Fujimori} one should note  very similar level of their quantitative conclusions. 
Despite the involvement of powerful calculation techniques the numerical results of the both approaches seem to be more like semiquantitative ones.  In such a situation we should transfer the center of gravity of the cluster approaches more and more to elaboration of physically sound and clear semiquantitative models that are maximally take into account all the symmetry  requirements  on one hand and refer to experiment on the other.

Hereafter, we do present a most recent and most comprehensive such a  cluster model approach to the description of the $p$-$d$ and $d$-$d$ CT transitions in NiO\,\cite{Sokolov} that nicely illustrates great potential of the model that does combine simple physically clear ligand-field analysis,  its semiquantitative predictions with a regular appeal to experimental data. We believe that such an approach should precede and accompany any detailed numerical calculation providing its physical validation. 
 
 Starting with an octahedral NiO$_6$ complex with  the point symmetry group $O_h$ we deal with  five Ni\,3$d$and eighteen  oxygen O\,2$p$ atomic
orbitals  
forming both the hybrid Ni\,3$d$-O 2$p$  bonding and antibonding $e_g$ and $t_{2g}$
molecular orbitals (MO), and the purely oxygen nonbonding $a_{1g}(\sigma)$, $t_{1g}(\pi)$,
$t_{1u}(\sigma)$, $t_{1u}(\pi)$, $t_{2u}(\pi)$ orbitals.
 The nonbonding $t_{1u}(\sigma)$ and $t_{1u}(\pi)$ orbitals with the same symmetry  are hybridized due to the oxygen-oxygen O 2\emph{p}$\pi$ - O 2\emph{p}$\pi$ transfer. The relative energy position of different nonbonding oxygen orbitals is of primary importance for the spectroscopy of the oxygen-3\emph{d}-metal charge transfer. This is firstly determined by the bare energy separation $\Delta \epsilon _{2p\pi \sigma}=\epsilon _{2p\pi }-\epsilon _{2p\sigma}$ between O 2$p$$\pi$ and O 2$p$$\sigma$ electrons.   Since the O 2\emph{p}$\sigma$ orbital points towards the two neighboring positive    $3d$ ions, an electron in this orbital has its energy lowered by the Madelung potential as compared with the O 2\emph{p}$\pi$ orbitals,    which are oriented perpendicular    to the respective 3\emph{d}-O-3\emph{d} axes. Thus, the Coulomb arguments  favor     the positive sign of the $\pi -\sigma$ separation      $\epsilon _{p\pi}-\epsilon _{p\sigma}$ whose   numerical value   can be easily    estimated in the frames of the well-known point charge model, and appears to be of the order of    $1.0$ eV.    In a first approximation, all the $\gamma (\pi )$ states     $t_{1g}(\pi),t_{1u}(\pi),t_{2u}(\pi)$ have the same energy. However, the O 2\emph{p}$\pi$-O 2\emph{p}$\pi$ transfer and overlap yield the energy correction    to the bare energies with the largest value and a positive  sign  for    the $t_{1g}(\pi)$ state. The energy of the $t_{1u}(\pi)$ state drops due to    a hybridization with the cation 4$p$$t_{1u}(\pi)$ state. 
   
   The ground state of NiO$_6$$^{10-}$ cluster, or nominally Ni$^{2+}$ ion corresponds to $t_{2g}^6e_g^2$ configuration with the Hund ${}^3A_{2g}(F)$ ground term.
    Typically for the octahedral MeO$_6$ clusters\,\cite{Moskvin-CT} the nonbonding $t_{1g}(\pi)$ oxygen orbital has the
highest energy and forms the first electron removal oxygen state while the other nonbonding oxygen $\pi$-orbitals, $t_{2u}(\pi)$, $t_{1u}(\pi)$, and the $\sigma$-orbital $t_{1u}(\sigma)$ have a lower energy with the energy separation $\sim$\,1\,eV inbetween (see Fig.\,1). 
   

\begin{figure}[h]
\includegraphics[width=8.5cm,angle=0]{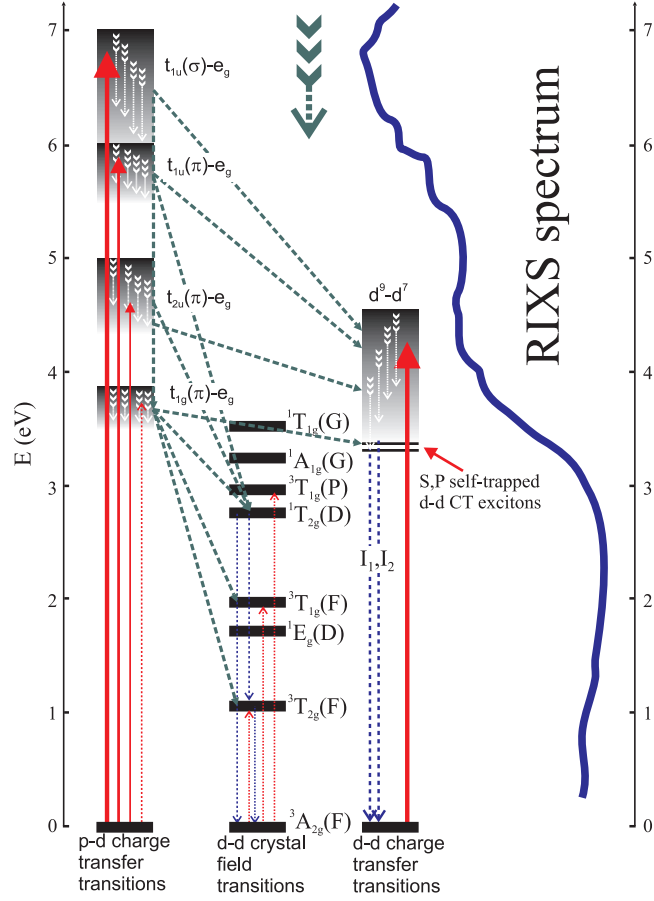}
\caption{(Color online) Spectra of the intersite \emph{d}-\emph{d}, \emph{p}-\emph{d} CT transitions and on-site crystal field \emph{d}-\emph{d} transitions in NiO. Strong dipole-allowed  $\sigma -\sigma$ \emph{d}-\emph{d} and \emph{p}-\emph{d} CT transitions are shown  by thick solid
uparrows;  weak dipole-allowed  $\pi -\sigma$ \emph{p}-\emph{d} transitions by thin solid
uparrows; weak dipole-forbidden low-energy transitions by thin dashed uparrows,
respectively. Dashed downarrows point to different electron-hole relaxation channels, dotted downarrows point to photoluminescence (PL) transitions, $I_{1,2}$ are doublet of very narrow lines associated with the recombination of the d-d CT exciton. The spectrum of the  crystal field \emph{d}-\emph{d} transitions is reproduced from Ref.\,\cite{Chrenko}. The right hand side reproduces a fragment of the RIXS spectra for NiO\,\cite{Duda}.} \label{fig1}
\end{figure}


The \emph{p}-\emph{d} CT transition in NiO$_{6}^{10-}$ center  is related to the transfer of O 2$p$ electron to the partially filled
3d$e_g$-subshell with the formation on the Ni-site of the $(t_{2g}^{6}e_{g}^{3})$ configuration of nominal Ni$^{+}$ ion isoelectronic to the well-known Jahn-Teller Cu$^{2+}$ ion. Yet actually instead of a single \emph{p}-\emph{d} CT transition we arrive at a series of O 2$p$$\gamma $$\rightarrow$ Ni 3$d$$e_g$ CT transitions forming a complex \emph{p}-\emph{d} CT band. 
It should be noted that each single electron $\gamma$$\rightarrow$$e_g$ \emph{p}-\emph{d} CT  transition starting with the oxygen $\gamma$-orbital gives rise to several many-electron CT states. For $\gamma$=$t_{1,2}$ these are the singlet and triplet terms ${}^{1,3}T_1$, ${}^{1,3}T_2$ for the configurations $t_{2g}^6e_g^3\underline{t}_{1,2}$, where $\underline{t}_{1,2}$ denotes the oxygen hole.
The complex \emph{p}-\emph{d} CT band starts with the dipole-forbidden $t_{1g}(\pi )$$\rightarrow$$e_g$, or ${}^{3}A_{2g}$$\rightarrow$${}^{1,3}T_{1g},{}^{1,3}T_{2g}$ transitions, then includes two formally dipole-allowed the so-called $\pi $$\rightarrow$$\sigma$ \emph{p}-\emph{d} CT transitions, the weak $t_{2u}(\pi )$$\rightarrow$$e_g$, and  relatively strong $t_{1u}(\pi )$$\rightarrow$$e_g$ CT transitions, respectively, each giving rise to ${}^{3}A_{2g}$$\rightarrow$${}^{3}T_{2u}$ transitions. Finally the main \emph{p}-\emph{d} CT band is ended by 
the  strongest dipole-allowed $\sigma $$\rightarrow$$\sigma$   $t_{1u}(\sigma )$$\rightarrow$ $e_g$ (${}^{3}A_{2g}$$\rightarrow$${}^{3}T_{2u}$) CT transition. The above estimates predict the separation between the partial \emph{p}-\emph{d} bands to be $\sim$\,1\,eV. Thus, if the most intensive CT band with a maximum around 7\,eV observed in the RIXS spectra\,\cite{Hiraoka,RIXS-2002,Duda} to attribute to the  strongest dipole-allowed  O 2$p$$t_{1u}(\sigma )$$\rightarrow$Ni 3d$e_g$ CT transition then one should expect the low-energy \emph{p}-\emph{d} CT counterparts with the maxima around 4, 5,  and 6\,eV\, respectively, which are related to the dipole-forbidden $t_{1g}(\pi )$$\rightarrow$$e_g$, the weak dipole-allowed  $t_{2u}(\pi )$$\rightarrow$$e_g$, and  relatively strong dipole-allowed  $t_{1u}(\pi )$$\rightarrow$$e_g$ CT transitions, respectively (see Fig.\,1). It is worth noting that  the $\pi $$\rightarrow$$\sigma$ \emph{p}-\emph{d} CT $t_{1u}(\pi)-e_{g}$
transition borrows a portion of the intensity from the strongest dipole-allowed $\sigma $$\rightarrow$$\sigma$ $t_{1u}(\sigma )$$\rightarrow$$e_g$ CT transition
because the $t_{1u}(\pi )$ and $t_{1u}(\sigma )$ states of the same symmetry are partly hybridized due to the \emph{p}-\emph{p} covalency and overlap.

Thus, the overall width of the \emph{p}-\emph{d} CT bands with the final  $t_{2g}^{6}e_{g}^3$  configuration occupies a spectral range from  $\sim 4$  up to $\sim 7$ eV. The left hand side of  Fig.\,1 summarizes the main semiquantitative results of the cluster model predictions for the energy and relative intensities of the \emph{p}-\emph{d} CT transitions.  
Interestingly this assignment finds a strong support in the reflectance (4.9, 6.1, and 7.2\,eV for the allowed \emph{p}-\emph{d} CT transitions) spectra of NiO\,\cite{Powell}. A rather strong $p$($\pi$)-$d$ CT band peaked at 6.3\,eV is clearly visible in the absorption spectra of MgO:Ni\,\cite{Blazey}.  The electroreflectance spectra\,\cite{electroreflectance} which detect the dipole-forbidden transitions clearly point to a low-energy forbidden transition peaked near 3.7\,eV missed in the reflectance and absorption spectra\,\cite{Powell,Blazey,Chrenko}, which thus  defines a \emph{p}-\emph{d} character of the optical CT gap and can be related to the onset transition for the whole complex \emph{p}-\emph{d} CT band. It should be noted that a peak near 3.8\,eV has been also observed  in the nonlinear absorption spectra of NiO\,\cite{Pisarev}.
At variance with the bulk NiO a clearly visible intensive CT peak near 3.6-3.7\,eV has been observed in the absorption spectra of NiO nanoparticles\,\cite{Volkov}. This strongly supports the conclusion that the 3.7\,eV band is related to the bulk-forbidden CT transition which becomes the partially allowed one in the nanocrystalline state\,\cite{Sokolov}.
It is worth noting that the hole-type photoconductivity threshold in bulk NiO has been observed  also at this "magic"\, energy	 3.7\,eV\,\cite{photoconductivity}, that is the $t_{1g}(\pi )$$\rightarrow$$e_g$ \emph{p}-\emph{d} CT transition is believed to produce itinerant holes.
Indeed, the $p$-$d$ CT transitions in NiO$_6$ cluster are of so-called "anti-Jahn-Teller"\,type, that is these are transitions from orbitally nondegenerate state to the final $p$-$d$ CT state state formed by two orbitally degenerate states that points to strong electron-lattice effects in excited state. The final Ni$^{1+}$  3$d$$^9$($t_{2g}^{6}e_{g}^3$)  configuration is isoelectronic to Cu$^{2+}$ ion in cubic crystal field and presents a well-known textbook example of a Jahn-Teller center that implies  a strong trend to the localization, while a photo-generated hole can move more or less itinerantly in the O 2$p$ valence band determining the hole-like photoconductivity\,\cite{photoconductivity}. It should be noted that any oxygen $\pi$-holes have a larger effective mass than the $\sigma$-holes, that results in a different role of the $p(\pi$)-$d$ and $p(\sigma$)-$d$ CT transitions both in photoconductivity and, probably, the luminescence stimulation.

A spectral feature near 6\,eV, clearly visible in the NiO photoluminescence excitation (PLE) spectra\,\cite{Sokolov}  can be certainly attributed to a rather strong  $p(\pi$)-$d$ ($t_{1u}(\pi )\rightarrow e_g$) CT transition while the spectral feature near 5\,eV to a weaker $p(\pi$)-$d$ ($t_{2u}(\pi )\rightarrow e_g$) CT transition. Interestingly the strongest $p(\sigma$)-$d$ ($t_{1u}(\sigma )\rightarrow e_g$) CT transition at $\sim$\,7\,eV is actually inactive in the PLE spectra, most likely, due to a dominating nonradiative relaxation channel for the oxygen  $t_{1u}(\sigma )$ holes.

However, the \emph{p}-\emph{d} CT model cannot explain the main low-energy spectral feature, clearly visible in the PLE spectra near 4\,eV\,\cite{Sokolov}, thus pointing to manifestation of another CT-type mechanism. Indeed, along with the \emph{p}-\emph{d} CT transitions an important contribution to the optical response of the strongly correlated 3$d$oxides can be related to the strong dipole-allowed \emph{d}-\emph{d} CT, or Mott transitions\,\cite{Moskvin-CT}.
In NiO one expects a strong  \emph{d}-\emph{d} CT transition related to the $\sigma -\sigma$-type  $e_g-e_g$ charge transfer $t_{2g}^6e_g^2+t_{2g}^6e_g^2$$\rightarrow$ $t_{2g}^6e_g^3+t_{2g}^6e_g^1$ between $nnn$ Ni sites with the creation of electron  NiO$_{6}$$^{11-}$ and hole NiO$_{6}$$^{9-}$ centers (nominally Ni$^{+}$ and Ni$^{3+}$ ions, respectively) thus forming a bound electron-hole dimer, or \emph{d}-\emph{d} CT exciton.

The strong dipole-allowed Franck-Condon $d(e_g)$-$d(e_g)$ CT transition in NiO manifests itself as a strong spectral feature near 4.5\,eV clearly visible in  the absorption of thin NiO films\,\cite{Rossi}, RIXS spectra\,\cite{Hiraoka,Duda}, the reflectance spectra (4.3\,eV)\,\cite{Powell}. Such a strong absorption near 4.5\,eV is beyond the predictions of the \emph{p}-\emph{d} CT model and indeed is lacking in the absorption spectra of MgO:Ni\,\cite{Blazey}. It should be noticed that, unlike all the above mentioned structureless spectra, the nonlinear absorption spectra\,\cite{Pisarev} of NiO films  do reveal an anticipated "fine"\, structure of the \emph{d}-\emph{d} CT exciton with the two narrow peaks at 4.075 and 4.33\,eV preceding a strong absorption above 4.575\,eV. Interestingly the separation 0.2-0.3\,eV between the peaks is typical for the exchange induced splittings in NiO (see, e.g., the "0.24\,eV"\, optical feature\,\cite{Chrenko}). 
Accordingly, the 4.1\,eV peak in the PLE spectra can be unambiguously assigned to the \emph{d}-\emph{d} CT transition\,\cite{Sokolov}. 

The charge, spin, and orbital degeneracy of the final state of this unique double anti-Jahn-Teller transition ${}^{3}A_{2g}+{}^{3}A_{2g}$$\rightarrow$${}^{2}E_{g}+{}^{2}E_{g}$ results in a complex band observed at 4.2-4.5 eV\,\cite{Sokolov}. The exchange tunnel reaction  Ni$^{+}$+Ni$^{3+}$$\leftrightarrow$Ni$^{3+}$+Ni$^{+}$  due to a two-electron transfer
gives rise to the two symmetric (S- and P-) excitons having s- and p-symmetry, respectively, with the energy separation $\delta_0=2|t|$ and $\delta_1=\frac{2}{3}|t|$ for the spin singlet and spin triplet excitons, where $t$ is the two-electron transfer integral whose magnitude is of the order of the Ni$^{2+}$-Ni$^{2+}$ exchange integral: $t\approx I_{nnn}$. Interestingly the  P-exciton is dipole-allowed while the S-exciton is dipole-forbidden. The anti-Jahn-Teller $d$-$d$ CT exciton is prone to be self-trapped in the lattice due to the electron-hole attraction and a particularly strong double Jahn-Teller effect for both the electron and hole centers. Recombination transitions for such excitons produce a bulk luminescence with puzzling well isolated doublet of very narrow lines with close energies near 3.3\,eV\,\cite{Sokolov} that corresponds to a reasonable Stokes shift of ~1 eV. To the best of our knowledge it is the first observation of the self-trapping for the $d$-$d$ CT excitons. 

Thus, we see that a simple cluster model is able to provide a semiquantitative description of a large body of experimental spectroscopic data, including subtle effects beyond the reach of any "ab initio" DFT technique. We have shown that the prototype 3$d$oxide NiO, similar to perovskite manganites RMnO$_3$ or parent cuprates such as La$_2$CuO$_4$\,\cite{Moskvin-CT}, should rather be sorted neither into the CT insulator nor the Mott-Hubbard insulator in the Zaanen-Sawatzky-Allen scheme\,\cite{ZSA}.

\section{Summary}

There are still a lot of people who think  the Hohenberg-Kohn-Sham DFT within the LDA has provided a very successful {\it ab initio} framework to successfully tackle the problem of the electronic structure of materials. However, both the starting point and realizations of the DFT approach have  raised serious questions. The HK "theorem" of the existence of a mythical universal density functional that can resolve everything looks like a way into Neverland, the DFT heaven is probably unattainable. Various DFAs, density functional approximations, local or nonlocal, will never be exact. Users are willing to pay this price for simplicity, efficacy, and speed, combined with useful (but not yet chemical or physical) accuracy\,\cite{Burke,Becke}.

The most popular DFA fail for the most interesting systems, such as strongly correlated oxides. The standard approximations over-delocalize the $d$-electrons, leading to highly incorrect descriptions. Some practical schemes, in particular,  DMFT can correct some of these difficulties, but none has yet become a universal tool of known performance for such systems\,\cite{Burke}. 

Any comprehensive physically valid description of the electron and optical spectra for strongly correlated systems, as we suggest, should  combine simple physically clear cluster ligand-field analysis with a numerical calculation technique such as LDA+MLFT\,\cite{Haverkort},  and a regular appeal to experimental data.

The research was supported by the  Ministry of Education and Science of the Russian Federation, project FEUZ-2020-0054.


\begin{thebibliography}{99}

\bibitem{HK}
H. Hohenberg and W. Kohn, Phys. Rev. {\bf 136}, B864 (1964).

\bibitem{KS}
W. Kohn and J.L. Sham, Phys. Rev. {\bf 140}, A1133 (1965).

\bibitem{Kryachko}
E.S. Kryachko, E.V. Ludena, Physics Reports 544 (2014) 123239.

\bibitem{Becke}
A.D. Becke,  J. Chem. Phys. {\bf 140}, 18A301 (2014).

\bibitem{LDA+U+DMFT}
G. Kotliar, S.Y. Savrasov, K. Haule, V.S. Oudovenko, O. Parcollet, and C.A. Marianetti, Rev. Mod. Phys. {\bf 78}, 865 (2006);
V.I. Anisimov and Yu.A. Izyumov, Electronic Structure of Strongly Correlated Materials (Springer Verlag, Berlin, 2010).


\bibitem{Eskes}
H. Eskes,  L.H. Tjeng, and  G.A. Sawatzky,  Phys. Rev. B {\bf 41},
288 (1990).

\bibitem{Ghijsen}
J. Ghijsen, L.H. Tjeng, J. van Elp, H. Eskes, J. Westerink,  G.A. Sawatzky, and M.T. Czyzyk, Phys. Rev. B {\bf  38}, 11322 (1988).


\bibitem{A}
L.H. Thomas, Proc. Cambr. Phil. Soc. {\bf 23}, 542 (1926).


\bibitem{B}
E. Fermi, Rend. Ac. Lincei {\bf 6}, 602 (1927).


\bibitem{Bobrov-JETP}
V.B. Bobrov, S.A. Trigger, JETP {\bf 143} , 729 (2013). 

\bibitem{Sarry}
A.M. Sarry, M.F. Sarry, Phys. Solid State, {\bf 54}(6), 1315 (2012). 


\bibitem{Ludena}
Eduardo V. Ludena,  J. Mol. Structure (Theochem) {\bf 709}  25 (2004).

\bibitem{Burke}
K. Burke,  J. Chem. Phys. {\bf 136}, 150901 (2012).

\bibitem{Amusia}
M.Ya. Amusia, A.Z. Msezane, V.R. Shaginyan, D. Sokolovski, Phys. Lett. A330:1-2 (2004), 10-15.


\bibitem{Sandratskii}
L.M. Sandratskii, Phys. Rev. B {\bf 64}, 134402  (2001).

\bibitem{Mazurenko}  
V.V. Mazurenko and V.I. Anisimov, Phys. Rev. B {\bf 71}, 184434 (2005); M.I. Katsnelson, Y.O. Kvashnin, V.V. Mazurenko, and A.I. Lichtenstein, Phys. Rev. B {\bf 82}, 100403(R) (2010).

\bibitem{MF}
A.S. Moskvin and S.-L. Drechsler, Phys. Rev. B {\bf 78}, 024102 (2008); Eur. Phys. J. B {\bf 71}, 331 (2009); A.S. Moskvin, Yu.D. Panov, S.-L. Drechsler, Phys. Rev. B {\bf 79}, 104112 (2009).
 


\bibitem{DC}
M. Karolak, G. Ulm, T. O. Wehling, V. Mazurenko, A. Poteryaev, and A. I. Lichtenstein, J. Electron Spectrosc. Relat. Phenom. {\bf 181},  11 (2010).

\bibitem{Nekrasov-JETP}
I.A. Nekrasov, N.S. Pavlov, and M.V. Sadovskii, JETP Lett. {\bf 95}(11), 581 (2012); JETP {\bf 116}(4), 620 (2013).

\bibitem{V} 
H.J. Kulik and N. Marzari, J. Chem. Phys. {\bf 134}, 094103 (2011); 
P. Hansmann, N. Parragh, A. Toschi, G. Sangiovanni, and K. Held, arXiv:1312.2757v1, 2013.

\bibitem{Brandow}
B. Brandow, Adv. Phys. {\bf 26}, 651 (1977); S. H\"{u}fner, Adv. Phys. {\bf 43}, 183 (1994).

\bibitem{ZSA}
J. Zaanen, G.\,A. Sawatzky, and J\,W. Allen, Phys. Rev. Lett. {\bf
55}, 418 (1985).

\bibitem{gap1}
G.A. Sawatzky,  J.W. Allen, Phys. Rev. Lett. {\bf 53}, 2239 (1984).

\bibitem{gap2}
S. H\"{u}fner, P.  Steiner,  I.  Sander,  F. Reinert,  and H.  Schmitt, Z. Phys. B - Condensed Matter 86, 207-215  (1992).

\bibitem{gap3}
E. Z. Kurmaev, R. G. Wilks, A. Moewes, L. D. Finkelstein, S. N. Shamin, and Kune$\check{s}$, PRB 77, 165127 (2008).

\bibitem{Terakura}
K. Terakura, T. Oguchi, A.R. Williams, and J. Kubler, Phys. Rev. B {\bf 30}, 4734 (1984).

\bibitem{GW}
S. Massidda, A. Continenza, M. Posternak, and A. Baldereschi, Phys. Rev. B {\bf 55}, 13494 (1997).

\bibitem{Anisimov_1993}
V.I. Anisimov, I.V. Solovyev, M.A. Korotin, M.T. Czyzyk, and G.A. Sawatzky, Phys. Rev. B {\bf 48}, 16929 (1993).



\bibitem{Nekrasov_2006}
X. Ren, I. Leonov, G. Keller, M. Kollar, I. Nekrasov, and D. Vollhardt, Phys. Rev. B {\bf 74}, 195114 (2006).


\bibitem{Anisimov_2007}
J. Kunes, V.I. Anisimov, A.V. Lukoyanov, and D. Vollhardt, Phys.Rev. B {\bf 75}, 165115 (2007).

\bibitem{NiO_optics}
C. Roedl and F. Bechstedt, Phys. Rev. B {\bf 86}, 235122 (2012).


 


\bibitem{Misha+Sasha}
M.I. Katsnelson, A.I. Lichtenstein, J. Phys. Cond. Matter. {\bf 22}, 382201 (2010).

\bibitem{Moskvin-CT}
A.\,S. Moskvin, A.\,V. Zenkov, E.\,I. Yuryeva, V.\,A. Gubanov,
Physica B, {\bf168},186 (1991); A.\,S. Moskvin, A.\,V. Zenkov,
E.\,A. Ganshina, G.\,S. Krinchik, and M.\,M. Nishanova, J. Phys.
Chem. Solids, {\bf 54}, 101 (1993); A.V. Zenkov,  A.S. Moskvin, J.Phys.Cond. Matter  {\bf 14}   6957 (2002); A.\,S. Moskvin, Phys. Rev. B {\bf 65},  205113 (2002); A.S. Moskvin, R. Neudert, M. Knupfer, J. Fink, and R. Hayn, Phys. Rev. B {\bf 65}, 180512(R) (2002); A.S. Moskvin, J. M\'{a}lek, M. Knupfer, R. Neudert, J. Fink, R. Hayn, S.-L. Drechsler, N. Motoyama, H. Eisaki, and S. Uchida, Phys. Rev. Lett. {\bf 91}, 037001 (2003); R.\,V. Pisarev, A.\,S. Moskvin, A.\,M. Kalashnikova, A.\,A. Bush,
and Th. Rasing, Phys. Rev. B {\bf 74}, 132509  (2006); A.\,S. Moskvin, R.\,V. Pisarev, Phys. Rev. B {\bf 77}, 060102(R) (2008); A.S. Moskvin, R.V. Pisarev, Low Temp. Phys. {\bf 36} 613 (2010); A.\,S. Moskvin, A.\,A. Makhnev, L.\,V. Nomerovannaya, N.\,N.
Loshkareva, and A.\,M. Balbashov, Phys. Rev. B {\bf 82}, 035106 (2010); A.\,S. Moskvin, Optics and Spectroscopy, {\bf 111}, 403 (2011).


\bibitem{Sugano}
S. Sugano, Y. Tanabe and H. Kamimura, Multiplets of Transition-Metal Ions in Crystals (Academic, New York, 1970).

\bibitem{Haverkort}
M.W. Haverkort, M. Zwierzycki, and O.K. Andersen, Phys. Rev. B {\bf 85}, 165113 (2012).

\bibitem{Fujimori}
A. Fujimori and F. Minami, Phys. Rev. B {\bf 30}, 957 (1984).
 
\bibitem{Mattheiss}
L.F. Mattheiss, Phys. Rev. B {\bf 5}, 290 (1972).

\bibitem{Sokolov}
V.I. Sokolov, V.A. Pustovarov, V.N. Churmanov {\it et al.}, JETP Letters, {\bf 95}, 528 (2012); Phys. Rev. B {\bf 86}, 115128 (2012); V.I. Sokolov, V.A. Pustovarov, V.N. Churmanov {\it et al.}, IOP Conf. Series: Materials Science and Engineering {\bf 38},  012007 (2012).

\bibitem{Hiraoka}
N. Hiraoka, H. Okamura, H. Ishii, I. Jarrige, K.D. Tsuei, and Y.Q. Cai, Eur. Phys. J. B {\bf 70}, 157 (2009).


\bibitem{RIXS-2002}
M. Magnuson, S.M. Butorin, A. Agui, and J. Nordgren, J. Phys.: Condens. Matter {\bf 14}, 3669 (2002).

\bibitem{Duda}
L.-C. Duda, T. Schmitt, M. Magnuson, J. Forsberg, A. Olsson, J. Nordgren, K. Okada   and A. Kotani,   Phys. Rev. Lett. {\bf 96}, 067402 (2006); B.C. Larson, Wei Ku, J.Z. Tischler, Chi-Cheng Lee, O.D. Restrepo, A.G. Eguiluz, P. Zschack, and K.D. Finkelstein, Phys. Rev. Lett. {\bf 99}, 026401 (2007).

\bibitem{Powell}
R.J. Powell and W.E. Spicer, Phys. Rev. B {\bf 2}, 2182 (1970).

\bibitem{Blazey}
K.W. Blazey, Physica  {\bf 89B}, 47 (1977).

\bibitem{electroreflectance}
J.L. McNatt, Phys. Rev. Lett. {\bf 23}, 915 (1969); R. Glosser, W.C. Walker, Solid State Commun, {\bf 9}, 1599 (1971).

\bibitem{Chrenko}
R. Newman and R.M. Chrenko, Phys. Rev. {\bf 114}, 1507 (1959).

\bibitem{Pisarev}
S.I.~Shablaev, and R.V.~Pisarev, Physics of the Solid State {\bf 45}, 1742 (2003).

\bibitem{Volkov}   
V.V. Volkov, Z.L. Wang, and  B.S. Zou, Chemical Phys. Lett. {\bf 337}, 117 (2001).


\bibitem{photoconductivity}
Ja.M. Ksendzov and I.A. Drabkin, Sov. Phys. Solid State, {\bf 7}, 1519 (1965).

\bibitem{Rossi}
C.E. Rossi  and W. Paul, J.  Phys.  Chem.  Solids {\bf 30}, 2295 (1969). 


\end{thebibliography}
\end{document}